\documentclass[conference]{IEEEtran}

\usepackage{amsmath,mathtools}
\usepackage{subfigure}
\usepackage{color}
\usepackage{algpseudocode}
\usepackage{algorithm}
\usepackage{amssymb}
\usepackage{ulem}
\usepackage{url}

\ifCLASSINFOpdf
\usepackage[pdftex]{graphicx}
\DeclareGraphicsExtensions{.pdf,.jpeg,.png} \else
\usepackage[dvips]{graphicx}
\usepackage{comment}
\fi

\usepackage[nolist]{acronym} 
\begin{acronym}
\acro{C-LOR}{co-location object relationship}  
\acro{C-WOR}{co-work object relationship}
\acro{DOSN}{decentralized online social network} 
\acro{ICT}{Information and Communication Technologies}
\acro{IoT}{Internet of Things}
\acro{IRN}{interested reachable node}
\acro{Online Social Network)}{OSN}
\acro{OOR}{ownership object relationship}
\acro{P2P}{peer-to-peer}
\acro{POD}{personal online datastore}
\acro{POR}{parental object relationship}
\acro{SIoT}{Social \ac{IoT}}
\acro{SOR}{social object relationship }  
\acro{URI}{Uniform Resource Identifier}
\end{acronym} 

\begin{document}

\title{Bridging Separate Communities with Common Interest in Distributed Social Networks through the Use of Social Objects}

\author{
\IEEEauthorblockN{
D.~Garompolo \textsuperscript{1}, A.~Molinaro \textsuperscript{1}, A.~Iera \textsuperscript{2}\\}
\IEEEauthorblockA{
\textsuperscript{1} Dept. DIIES, University ``Mediterranea'' of Reggio Calabria\\
\textsuperscript{2} Dept. DIMES, University of Calabria, Arcavacata di Rende (CS), Italy\\
email: [david.garompolo, antmolin]@unirc.it, antonio.iera@unical.it\\
}
}

\maketitle

\begin{abstract}
In light of the growing number of user privacy violations in centralized social networks, the need to define effective platforms for decentralized online social networks (DOSNs) is  deeply felt. Interesting solutions have  been proposed in the past, which own the necessary mechanisms to allow users keeping control over their personal information and setting the rules to regulate the access of other users. Unfortunately, the effectiveness of this type of solutions is severely reduced by the fact that different user communities with a shared interest could be disconnected/separated from each other.  This translates into a reduced ability in effectively spreading data of common interest towards all interested  users, as it currently happens in centralized social networks.
In order to overcome the cited limitation, this paper proposes a disruptive approach, which exploits the availability of a new class of Internet of Things (IoT) devices with autonomous social behaviors and cognitive abilities. Such devices can be leveraged as friendship intermediaries between devices' owners who are connected to a DOSN platform and share the same interest.
We will demonstrate that clear advantages can be achieved in terms of increased percentage of  Interested Reachable Nodes (a specific measure of Delivery Ratio) in distributed social networks among humans, when enhanced with so called \textit{Mediator Objects} adhering to the well-known social IoT (SIoT) paradigm.
\end{abstract}


\IEEEpeerreviewmaketitle

\section{Introduction}
\label{sec:1}

Recent scandals such as the one involving Cambridge Analytica have clearly shown that technology giants (Facebook, Google \& Co, etc.) have serious problems in protecting user data. In order to guarantee users  privacy protection and a greater control over their data, researchers have recently proposed several distributed platforms to implement \acp{DOSN} \cite{guidi2018managing} either based on \ac{P2P} architectures \cite{aiello2010secure,buchegger2009peerson,cutillo2009safebook,guidi2016didusonet} or with a Web-based nature \cite{yeung2009decentralization,diaspora}. 

In this context, an interesting decentralized platform for social Web applications, named Solid, was proposed by Tim Berners Lee in  \cite {sambra2016solid}. It  decouples user data from applications that create and consume this data, ensuring they have a simple, generic and well-defined way to access data stored in the  user's Web-accessible \ac{POD} \cite{sambra2016solid}. This platform, in essence, allows users to maintain control over their personal information and to decide where it is stored, who can view it, and which applications can access it.

Web-oriented \acp{DOSN} can be, for example, developed by leveraging the Solid platform (Solid Social is an example) 
through which a user (\textit{Source}) can diffuse her own content in a targeted manner to all people interested in receiving it. In Solid, the social graph visible to each user consists of the contacts stored in her POD, the contacts of these contacts, and so on. Contacts can be seen as an interface to manage the user's distributed social graph. Interestingly, a user can either mark her contacts as public or make them  accessible to a specific individual or group of people \cite{sambra2016solid,mansour2016demonstration}. While with a centralized social network the application knows the complete social graph, in the distributed case it only knows a partial graph consisting of (i) the contacts stored in the \ac{POD} of the Source willing to disclose the content and (ii) the contacts in the \acp{POD}, which the Source is authorized to access.
Obviously, authorization mechanisms in the use of the node’s contacts can be leveraged to guarantee a given content to be diffused to the interested nodes only. This allows reducing as much as possible the number of unnecessary visited nodes  and  making the information spreading as efficient as possible 
\cite{gao2011user, mashhadi2009habit, kong2014semi, kong2016content}.

Unfortunately, such an approach, although being very effective with a view to ensuring trust and security in accessing the information, at the same time amplifies the inherent limitations of decentralized social network solutions: \textit{different communities with the same interest could very likely be disconnected/separated from each other, and some nodes potentially interested in a content could be isolated}. 
In other words, nodes that share the same generic interest cannot always be mutually discovered and cannot exchange contents related to that interest. This reduces the extent to which a Source is able to spread its content to the highest number of possibly interested nodes.

The associated risk is a decrease in the appeal of this type of solutions for the users playing the role of ``prosumers" (i.e., both producers and consumers) of contents. As consumers, they could see the access to the content of their interest threatened by a discovery process hindered by fragmentation and possible isolation of some communities of interest. Even worse is the situation for a user who produces contents. In such a case, the producer would have difficulty in spreading her content to a large number of interested users and, if a business is based on the delivery of such a content, the problem implies a reduction in revenues from commercial activities that operate on distributed  platforms.

A new class of devices, showing autonomous social behaviors and  having cognitive abilities, could help in addressing some of the highlighted issues. The idea is to allow such devices to act as ``facilitators" of contacts and friendship among users with the same interests and, therefore, behaving as ``bridges" between communities with similar interests which would not be connected in a distributed social network.

\textcolor{black}{The main contributions that our research intends to provide are summarized below:
\begin{itemize}
\item  the well-known Social Internet of Things (SIoT) paradigm \cite{atzori2012social}\cite{atzori2014smart} and the possibility that it offers to have a new class of devices, showing autonomous social behaviors and having cognitive abilities, is leveraged to increase the number of interested nodes that can be reached by a given content in a DOSNs based on the Solid platform;
\item  the new concept of \textit{Mediator Object} is introduced, which refers to a device with a social and cognitive nature that mediates the content propagation towards other interested devices/users, otherwise unreachable.  This concept could be seen as the first implementation, through Information and Communication Technologies (ICT), of the well-known concept of “social object” in sociological studies on object centered sociality \cite{cetina}, referring to an object creating a human connection between two individuals;
\item  the SIoT paradigm is evolved into an empowered version, named here \textit{Enhanced SIoT}. It still exploits the basic management and control mechanisms proposed in the literature for the SIoT (included those related to \ac{IoT} devices’ trustworthiness  \cite{chen2015trust}, \cite{chen2016scheme}, \cite{nitti2012subjective}), and adds new features to enhance the information diffusion among social devices, and thus to increase the number of interested nodes that can be reached by a given content (interested reachable nodes - IRNs); 
\item  a set of analysis are conducted that show the advantages offered in terms of increased percentage of IRNs in a decentralized social network, when leveraging the introduced paradigms of Object Mediators and Enhanced SIoT, compared to the traditional case in which only friendships among humans are considered; 
\item  a last additional side contribution is also the proposal of a possible methodology for modeling a SIoT network between devices starting from real human tracks in Social Networks, which can be useful to researchers in the field of SIoT, because today there are no tracks nor datasets from real SIoT devices available yet.
\end{itemize}}

This paper is organized as follows. In Section II background details on the Solid system and the SIoT paradigm are provided.
Section III summarizes the reference literature while Section IV addresses the proposed Enhanced SIoT, illustrates the concept of community and the role of the Mediator Object, and describes how discovery and diffusion take place in the Enhanced SIoT.
An analysis of the performance achieved by the proposed solution compared to a traditional centralized social network solution is the subject of Section V, while conclusive remarks are given in Section VI.

\section{Background}
\label{sec:2}

\subsection{Distributed vs. Centralized Online Social Networks}
\textcolor{black}{A Decentralized Online Social Network (DOSN) is an online social network implemented on a distributed platform. By distributed we mean that all computing, storage and communication resources are provided by the users.}

\textcolor{black}{\acp{DOSN} are considered in many works in the literature  as a possible solution to give users greater control over their data and, at the same time, to overcome typical problems of centralized online social networks, such as privacy, performance bottleneck, single point of failure, need to be online in all transactions, unexploited locality \cite{guidi2018managing, guidi2019towards, mega2018social, yuan2018efficient}.}

\textcolor{black}{While in the Centralized case the single service provider has control over all user data and could change the existing terms of service, in the Decentralized case there is a set of nodes that cooperate to guarantee all the functionalities. This gives to the users more control over their privacy \cite{Conti}.}

\textcolor{black}{Indeed, in a DOSN the user data will be stored locally in devices and controlled by end users instead of the OSN central entity (OSN provider). Consequently, the DOSNs can mitigate the privacy control issues, the problem of security and scalability and increase the flexibility and the ability to deal with big data problems \cite{yuan2018efficient} \cite{paul2011decentralized}.} 

\textcolor{black}{Another important problem of Centralized OSNs is the performance bottleneck due to the very high number of user requests and the huge amount of social data (all data exchanged in Social Networks regarding both user information and generated content). The DOSNs can alleviate the problem of performance bottleneck and avoid the single point of failure and the single point of attack. }

\textcolor{black}{Finally, centralized OSNs can suffer from other problems such as limited scalability and high maintenance costs to manage the data of so many users. In DOSNs, shifting to the user both the implementation of the infrastructure and the privacy and security control effectively reduces the operational cost \cite{Buchegger2},\cite{Graffi}.}

\textcolor{black}{Table I synthesizes the main differences between centralized and decentralized OSN
}


\begin{table*}[htbp]
\textcolor{black}{\caption{Centralized vs Distributed OSN}}
\centering
\setlength{\tabcolsep}{10pt}
\renewcommand{\arraystretch}{1.5}
\begin{tabular}{p{3cm}p{5cm}p{5cm}}
\hline

\textbf{  }	& \textbf{Distributed OSNs}   & \textbf{Centralized OSNs}\\
\hline

\textbf{Computing, storage and communication resources} & Provided by the users &   Provided by the OSN central entity (OSN provider) \\
\hline
\textbf{User data storage}  & Local data storage in devices controlled by end users	& Remote data storage controlled by the OSN central entity (OSN provider)\\
\hline
\textbf{Privacy and Terms of Service}  & Users have more control over their data privacy & The service provider has control over all user data and can change the terms of service\\
\hline
\textbf{Security, Scalability, and Maintenance costs to manage users data} & Mitigate security and scalability issues and increase the flexibility in dealing with big data problems & May suffer from security control and scalability issues 
 \\
 \textbf{   } &  Achieve an effective reduction of operational costs by transferring infrastructure, privacy and security control to users & May suffer from high maintenance costs to manage data from a huge number of users
 \\  
\hline
\textbf{Performance bottleneck} &  Counter the performance bottlenecks caused by high number of users requesting social data regarding both user information and generated content &  Exposed to severe performance bottlenecks having to centrally manage the huge amount of social data and user requests
\\
\hline
\textbf{Single Point of Failure \& Single Point of Attack} & Avoid single point of failure and single point of attack. related issues & May suffer from single point of failure and single point of attack issues
\\
\hline

\end{tabular}
\label{tab:actors}
\end{table*}


\subsection{Solid}

Proposed by Tim Berners-Lee, Solid is a web-based open source platform that allows users keeping control over their personal information. Thanks to it, large social network companies will be allowed to use only part of the user data, and this permission can also be withdrawn at any time. 

The Solid's core is represented by the Solid \ac{POD}, which can be seen as ``a private website with data inter-operable with all apps". It stores all the user's personal information that will be linked from the outside in order to be used. In this model, only  users will control their own information \cite{sambra2016solid, mansour2016demonstration}.
The ``Contacts" application manages a list of contacts stored on a user's \ac{POD}. In Solid the user's social graph consists of the contacts stored in the \ac{POD}, plus the accessible contacts of these contacts and so on; each user is identified by a WebID. The Contacts application maintains a set of vCards for the user's contacts by using the vCard ontology. Each vCard is a resource with a single \ac{URI} and can contain the user's WebID that it represents, in addition to other fields such as name and email. A user can mark a vCard as ``public" or allow access to the vCard only to a single individual or to a specific group of people (identified by their WebIDs).

This new vision, which allows decoupling data from applications, brings with it new issues to be addressed in view of deploying social network applications for Solid (e.g., Solid Social). While in the centralized social networks the application (e.g. Facebook, Instagram, etc.), possessing all the data (in the Data Silos), knows the complete graph of the contacts of all users, in this new distributed vision this is no longer possible. In particular, each user will know, in addition to her contacts contained in her Solid \ac{POD}, only the contacts contained in the Solid \acp{POD} of the users who have granted her permission to access their own \acp{POD}. Consequently, there will be a graph with many separate components (Local Knowledge). 

Let us imagine that a user wants to disclose a given content in a targeted manner to all the nodes that are interested in receiving it. In this case, being the social network application distributed and therefore no longer aware of the complete graph, during the discovery phase it might not be able to reach all the nodes interested in receiving the content, as it was in the centralized case. Consequently, even during the targeted diffusion phase (i.e., diffusion  to the interested nodes only) a node may not be able to spread its content to all the other interested nodes.


\subsection{\ac{SIoT}}
The \ac{IoT} is a paradigm that in recent years has received much attention from the scientific community. 
A possible way to look at the \ac{IoT} is to define it as \textit{``a conceptual framework that leverages the availability of heterogeneous devices and interconnect solutions, as well as augmented physical objects that provide a shared global information base to support the design of applications that involve both people and representations of objects"} \cite{atzori2017internet}. 
Accordingly, every person and thing in the \ac{IoT} has a virtual counterpart that can be localized, addressed, and readable in the Internet (in the Cloud, the Fog, or at the network Edge).

Objects act as prosumers of services and collaborate with other counterparts to reach common goals. This leads to design a new generation of ``smart objects" that will have to operate in complex contexts; it is unlikely that the single object will have the capacity to deal with such a complexity. Like some species of animals that, to cope with complexity and the difficulties of the environment in which they live, have created a dense network of social relationships, in the same way a new generation of social objects has been envisaged to form an augmented \ac{IoT}, called \ac{SIoT}, that applies to \ac{IoT} the typical concepts, solutions, and technologies of social networks \cite{schurgot2012beyond}. Accordingly, an \ac{IoT} object becomes part of and acts in a social community of objects and devices. In this context, the social networks of objects are built among objects that are owned by human beings, who may have no connection among them \cite{atzori2014smart}.

The application of the social networking principles to the \ac{IoT} (i.e., \ac{SIoT}) can bring several advantages in network navigability, scalability, and discovery of objects and services. Moreover, powerful models of trustworthiness management designed for social networks can be reused to address \ac{IoT}-related issues. 

In \ac{SIoT} a basic set of inter-device relationships are defined: \ac{POR}, established between objects belonging to the same production batch; \ac{C-LOR}, established between objects used in the same location; \ac{C-WOR}, established between objects that often cooperate together to provide a common IoT application; \ac{OOR}, established between objects  belonging to the same user, and \ac{SOR}, established between objects that often come into contact because their owners come into contact with each other during their lives.

The creation and management of such relationships can take place without human intervention \cite{atzori2019social}, but always strictly complying to rules set by the devices’ owners.  Several further types of friendship have been added to these relationships over the time, driven by specific application environments (e.g., the social Internet of vehicles).

\section{Related Works}
\label{sec:3}

\subsection{DOSN classification}

In \cite{paul2011decentralized} \acp{DOSN} are classified into two categories: Web-based and \ac{P2P}-based \acp{DOSN}. The first category, which is considered as a basis in our work, is characterized by a distributed web server infrastructure, and includes systems such as Diaspora \cite{diaspora} and ``Friend-of-a-Friend" (FoaF) \cite{yeung2009decentralization}. These solutions need web space or deploying web servers. Users can publish their profiles in their Web space and manage access control rules (access authorizations) locally, to specifically allow the recovery of attributes and resources reserved for the selected users. Web links to other users' profiles are used to represent the Contact List and thus recreate the social graph.

The second category includes systems such as Likir \cite{aiello2010secure}, Peerson \cite{buchegger2009peerson}, Safebook \cite{cutillo2009safebook}, which exploit the advantages of the \ac{P2P} principle  to allow the publication, search, and retrieval of profiles and their attributes, very similarly to conventional \ac{P2P} file-sharing systems. In them, the resources are kept locally and the profiles are stored in the local devices instead of the Web, and controlled by the users themselves. In \ac{P2P}-based \acp{DOSN} one of the main challenges concerns the replication of user profiles, which must always be available online even when the user is offline. The issue of data availability/persistence is addressed in \cite{guidi2018managing}. 

In summary, while web-based systems rely on a dedicated web space where user profiles can be stored and retrieved, \ac{P2P}-based systems exploit only the local and shared resources of the \ac{P2P} overlay. Obviously, exploiting the rather unreliable storage services of other peers, which are subject to churn (i.e., nodes entering and leaving the network, or changing state from online to offline continuously), requires more sophisticated means to keep the data available, which in turn causes overhead and higher implicit costs shared among the participants. Differently, in our research we follow a Solid-like web-based system design and, since the \acp{POD} in which the profiles are stored are always available online, there is no need for replicas of the same profile.

\subsection{Content sharing and diffusion}
Our reference scenario is characterized by the presence of Interest Communities 
(i.e., sets of nodes that share the same interest) that are separated, disconnected, and unreachable from each other. This represents the problem we aim to solve. 

The reasons for which the concerned nodes  can be unreachable are the most disparate and vary according to the considered scenario. They can be related \textit{(i)} to the overlay topology; for example, in Friend-to-friend (F2F) networks, communications can only take place between ``friends"; \textit{(ii)} to the type of content diffusion; for example, in the case of targeted diffusion, the content is sent only to the interested nodes without any bridging performed by non-interested intermediate nodes; or \textit{(iii)} to the fact that the interested nodes are located in different network Partitions (i.e.,  areas of a network that are connected in the presence of a mobile node and disconnected in its absence) caused by churn phenomena, for example.
This is not an exclusive feature of \acp{DOSN}, but it may also occur in other systems, ranging from mobile ad-hoc to opportunistic networks.

\textcolor{black}{Several methods have been proposed in the literature that consider the community structure to be a significant property of social networks and propose methods for community identification, most of which apply to static networks. There are also interesting studies available, as the one described in \cite{Zhixiao}, focusing on dynamic and overlapping community detection. There, the authors propose a novel method able to track the evolution of overlapping communities in dynamic social networks based on topology potential field, which jointly solves the problems of overlapping community detection, dynamic community identification and community evolution analysis. Through this method dynamic overlapping social networks can be accurately partitioned and all kinds of community evolution events efficiently tracked.}

In general, the goal of a content-sharing system is to move content items to devices owned by users who want to access such a content \cite{nordstrom2014haggle}. In our reference context the goal is very similar, as a Source node wants to spread its content to the highest possible percentage of  interested reachable nodes. The main issue in such a scenario is that, when diffusing the information relevant to the Source's interests both to the interested and not interested nodes during the Discovery phase, not all nodes will be equally cooperative and equally reachable. Also, they will not be equally willing to authorize the access to their PODs and to re-forward information on behalf of the Source to their Contacts, otherwise unreachable  by the source directly. 
The solutions proposed in the literature to spread contents to interested users belonging to different Communities of Interest/Partitions are manifold and usually depend on the type of scenario and related assumptions. 
In \cite{nordstrom2014haggle} an ad-hoc content-sharing system for mobile devices is proposed within an opportunistic network scenario. In it, mobile devices share contents and interests and can store-carry-forward contents on behalf of other nodes, based on interests, therefore connecting otherwise disconnected devices. In \cite{nordstrom2014haggle}, Haggle introduces a content delegation mechanism that allows to selflessly disseminate a given number of items based on the interests of other nodes (third-party nodes). This is particularly important in the presence of network Partitions. 
The mobile nodes that provide this type of connectivity are called ``data mules", reflecting the idea that they carry data between otherwise partitioned areas of the network, \cite{birrane2020designing}. In \cite{nordstrom2014haggle} it is clearly explained that depending on the network structure and the users' interests, a content may not reach the interested nodes without exploiting data mules that carry the content although they are not interested in it. Although the scenario is very different from ours, it presents a problem very similar to the one we intend to address. Here too, the goal is to spread the contents to devices owned by users who need such contents \cite{nordstrom2014haggle}. 


Several studies aimed to design valid mechanisms to diffuse the content as much as possible to the interested nodes. For example, \cite{kong2014semi} analyses how maximizing the total weight of the content receivers, which is measured in proportion both to how much the users themselves are interested in the content and based on their ability to connect with other interested users.

There are many reasons why nodes should decide to spread the information only to the interested nodes. First, the latter are more active during the diffusion process \cite{kong2016content}.
The second important reason is the guarantee of greater privacy and security. So a possible research goal is to minimize the number of nodes not interested that are involved in the diffusion process \cite{gao2011user}, \cite{mashhadi2009habit}.
In our case, even in the Discovery phase, when information on the Source interests is diffused, we are able to guarantee privacy by sending it totally anonymously. 

In \cite{boldrini2008contentplace} the goal is to make the information available in those regions where there are interested users without overusing the resources, i.e., avoiding flooding.
Third, it must be considered that the  interested nodes are also the desired receivers, and therefore they are motivated to participate in the diffusion process to receive the content of interest. 
Finally, users with similar interests also have a higher frequency and probability of communicating with each other, which allows a more efficient diffusion  \cite{kong2016content}. This feature is exploited to improve content diffusion in \cite{ciobanu2014onside}. In particular, a user will download the content from another user she meets if and only if \textit{(i)} the topic of the content is of her own interest, \textit{(ii)} it interests her friends, or \textit{(iii)} it interests the users she meets. 

In our framework the content of the Source is diffused  \textit{only to interested nodes} and, consequently, there will be no uninterested data mules that connect the separate Communities of Interest (Partitions).  Differently, during the Discovery  phase the information on the Source's interests will be sent (anonymized) \textit{both to interested and not interested nodes}. Intermediary nodes (even if not interested) spread the information relevant to the Source's interest  and allow the Source to reach nodes that otherwise it would not be able to directly reach.
As a result, on the one hand, during Discovery, the Source is able to indirectly reach  a higher number of interested nodes, thanks to the presence of intermediaries. On the other hand, during Diffusion, the Source manages to send its content as efficiently as possible to the interested nodes only. 

Privacy is guaranteed both during the Discovery phase, thanks to the anonymity of the information disseminated, and, during the Diffusion phase, by sending the content directly to the desired receivers only.

Due to the churn  
in the Social Overlay (SO), a limited set of links may be available for reconfiguration and cause transient network partitions, which are responsible of long unacceptable delays in the content diffusion phase \cite{mega2018social,mega2012churn}.
As a solution to this issue, a hybrid architecture is proposed in \cite{mega2018social} that on the one hand exploits the SO for fast, decentralized and friend-to-friend communications, but occasionally exploits the access to the cloud to overcome the high delays caused by the purely decentralized solution.
Comparing it with our work, we see the analogy between the transient network partitions and the separate Communities of Interest, but the focus is slightly different because, while in our case a Source node wants to spread a content to all the interested network nodes, the goal in \cite{mega2018social} is to allow efficient profile-based communication between direct friends. In both cases, however, the problem lies in the limited set of available links.
Unlike the solution presented in \cite{mega2018social}, in which the concept of purely distributed architecture is lost, in our case we want to show the advantages in terms of reachability that are obtained by extending the set of friendships through social object relationships (SIoT Contact List), while still maintaining a 
distributed approach to reach any interested node belonging to other communities.

\section{Enhanced Social Internet of Things}
\label{sec:4}
\subsection{Community}


Fig. \ref{fig:meccanismobase} shows two Interest Communities in our reference scenario.
Let us assume, for example, that Community $A$ is made up of nodes that share at least one common interest, e.g. the ``Football" Interest. This does not mean that nodes belonging to a community cannot have other interests in common with nodes belonging to other communities, but, to simplify, in Fig. \ref{fig:meccanismobase} we do not illustrate this case. 
\textcolor{black}{In the remainder of the section, let us assume that:}\\\\
\textcolor{black}{- $C_A$ = Community A = Community of nodes that share at least the "Football" Interest = $C(I_A)$;}\\
\textcolor{black}{- $I_A$ = Interest A = Interest "Football";}\\
\textcolor{black}{- $V(I_A)$ = set of all nodes (vertices of the graph) with $I_A$;} \\
\textcolor{black}{- $C_A$ $\subseteq$ $V(I_A)$. }\\

Each direct dashed line between a pair of nodes indicates that both nodes can be mutually reached by performing a Search for that specific interest. They can be reached using the ``Contacts of Contacts" mechanism, as they are authorized to access the contacts of their contacts. If two nodes within the same community are not connected by a dashed line, it means that they cannot be reached via direct search, i.e., if one of the two nodes searches for that interest, the other's vCard will not be returned.

\textcolor{black}{Taking up the concept of Reachability from Graph Theory we know that:
a node $n_2$ is reachable by a node $n_1$ (or $n_1$ reaches $n_2$) if there exists a path from $n_1$ to $n_2$.
We define two relations on the set of nodes:\\\\
(a) the reachability relation $R$$\to$ such that $n_1$$R$$\to$$n_2$ if $n_2$ is reachable from $n_1$;\\
(b) the mutual reachability relation $R$$\rightleftarrows$ such that $n_1$$R$$\rightleftarrows$$n_2$ if $n_1$$R$$\to$$n_2$ and $n_2$$R$$\to$$n_1$.}\\

\textcolor{black}{At this point, considering that in our scenario the mutual reachability relationship indicates that “\textit{both nodes can be mutually reached by performing a Search for that specific interest and they can be reached via the “Contacts of Contacts” mechanism, as they are authorized to access the contacts of their contacts}”, we can represent this relationship with the following formalism:\\ \\
$n_1$$R$$\rightleftarrows$$n_2$, which graphically, in Fig. \ref{fig:meccanismobase}, corresponds to the direct dashed line between $n_1$ and $n_2$, with $n_1$,$n_2$ $\in$ $C_A$}\\

A node of a community that intends to spread the content belonging to a specific interest is called Source. This node,  thanks to the permissions received from its contacts to view their contacts, can directly reach with its content the  interested nodes. 
The reachable nodes are defined as Direct Interested Reachable Contacts (D-IRC) of the Source.
Furthermore, its contacts may relaunch the search to reach further interested nodes (otherwise not reachable from the Source) by operating as Sources. This hopefully allows the Source's  content to reach all the nodes interested in that specific interest (the respective IRCs), which for the Source are to be considered Indirect IRCs (I-IRCs). 

\textcolor{black}{We must now distinguish two types of relationship:}
\begin{itemize}
\item 

\textcolor{black}{$DR$$\to$ = Direct Reachability}

\textcolor{black}{It holds if a node, thanks to the permissions received from its contacts to view their contacts, can directly reach with its content the interested nodes; in this case, 
the set D-IRC 
of the Source node $S$ is expressed as:}\\

\textcolor{black}{\textit{D-IRC(S)} = \{$\forall$ $n$ $\in$ $V(G)$ $|$ $S$ $DR$$\to$ $n$, $n$ $\subseteq$ $V(I_A)$\}}\\

\textcolor{black}{given that:} \\

\textcolor{black}{$d_A$ = $d(I_A)$ = data (content) characterized by Interest A;}

\textcolor{black}{$S$ = the Source that wants to disseminate $d_A$;}

\textcolor{black}{$S$ $DR$$\to$ $n$ $\in$ \textit{D-IRC(S)};}

\textcolor{black}{$V(G)$ = set of all nodes of the graph $G$;}\\

\textcolor{black}{and represents the set of nodes, directly reachable by $S$, interested in the content characterized by Interest A}

\item 
\textcolor{black}{$IR$$\to$ = Indirect Reachability}

\textcolor{black}{It holds if the node's contacts may relaunch the search to reach further interested nodes (otherwise not reachable from the Source) by operating as Sources. In this case} \textcolor{black}{The set \textit{I-IRC} 
of the Source node $S$ is expressed as:}\\

\textcolor{black}{\textit{I-IRC(S)} = \{$\forall$ $n$ $\in$ $V(G)$ $|$ $S$ $IR$$\to$ $n$, $n$ $\subseteq$ $V(I_A)$\} 
}\\

\textcolor{black}{given that:} \\

\textcolor{black}{$S$ = the Source that wants to disseminate $d_A$;} 

\textcolor{black}{$S$ $IR$$\to$ $n$ $\in$ \textit{I-IRC(S)};}\\

\textcolor{black}{and represents the set of nodes, indirectly reachable by $S$, interested in the content characterized by Interest A}
\end{itemize}

By applying this process recursively, a Source node will be able to disclose its content to all interested nodes of a community (both D-IRCs and I-IRCs);  specifically, all the reachable nodes will constitute the community. 

\textcolor{black}{Let  $S$ be the Source that wants to disseminate $d_A$.} \textcolor{black}{The set of nodes directly reachable from $S$ plus that of nodes indirectly reachable from $S$, interested in the content characterized by Interest A, is equal to the set of all the nodes reachable from $S$, interested in the content characterized by Interest A.}\\
\textcolor{black}{We call this set $C_A$ =  \textit{D-IRC(S) + I-IRC(S)} = $IRC(S)$,\\
defined as:}\\
\textcolor{black}{$C_A$ = \{$\forall$ $n$ $\in$ $V(G)$ $|$ $S$ $DR$$\to$ $n$ $||$ $S$ $IR$$\to$ $n$, $n$ $\subseteq$ $V(I_A)$ \}\\
}

Finally, we assume that if two nodes sharing the same interest are in two different communities, then none of the nodes in the first community can reach any node in the second community. In Fig. \ref{fig:meccanismobase} the node marked as ``interested node" is a node that cannot be reached by searching from any node belonging to Community $A$, otherwise it would belong to it, and therefore it is part of Community $B$.

In order to enable content diffusion also to interested nodes belonging to other separate communities, we propose here to exploit the SIoT concept, as explained in the following subsection.

\subsection{Mediator Object}
The idea of a Mediator Object is leveraged  in this context to reach otherwise unreachable interested nodes; its objective is precisely to ``mediate" the content propagation from Community $A$, wherein the content is generated by the Source node, to a separate Community $B$, where the interested node is located. This sort of ``bridging" between two communities can be achieved through objects that have a social and cognitive behavior and interact with one another. 

In Fig. \ref{fig:meccanismobase}, we consider a SIoT network between the two  communities  made up of cognitive objects. In the remainder of the paper, by \textit{"cognitive object"} we will indicate an object with the ability to proactively search the social network of objects to which it belongs (through the use of the SIoT platform) and to understand, from the Interest Descriptors (defined in the next Section) and from previous events, whether ``friend objects" from other communities may be interested in receiving a certain content. The way it happens will be described in the following. \textcolor{black}{Obviously, in order to carry out its functions of mediation between communities, the Mediator Object must be a cognitive object.}

An exemplary use case is given by a cognitive object which understands, through the mechanism described below, that the news circulating in Community $A$ of soccer fans may also interest another  ``friend" and ``trusted" object, whose owner belongs to Community $B$ of soccer bettors, for example. We assume that a social object relationship was indeed created between the devices, according to the SIoT rules. The devices became friends, because, for example, they often came into contact in a soccer stadium, although their owners do not know each other.

In this context, a first problem is linked to the way in which a cognitive object can know if ``friend objects" from other communities may be interested in receiving a certain content.
To this end, each device is associated with an Interest Descriptor, i.e. a vector of words (keywords) that describe the interests of its owner.

There are several ways to derive the Interest Descriptor, depending on the information available.
Without losing generality,  we can refer to an exemplary solution based on the Visual User Interest Profile (VUIP), as in \cite{zhou2016applying}. 

In order to better understand what the VUIP is, it is necessary to introduce the concept of user profiling, which can be defined as the process of identifying data relating to the user's domain of interest.
A device can infer the owner's profile based on a set of images, accessed  through that device, that describe her interests (as in Instagram, Facebook, etc.). For example, by leveraging deep learning techniques, from the images it is possible to obtain the corresponding VUIP (that is a vector of keywords derived from images) that can be used as an Interest Descriptor. 

\textcolor{black}{The method we will adopt is just one of several possibilities allowing to extract the keywords (tags) associated with images. As an example, another work that deals with Image Tagging problem, i.e. the extraction of tags from the image, is described in \cite{Zechao}. The authors describe a new Deep Collaborative Embedding (DCE) model for social image understanding applied to social image tag refinement and assignment, content-based image retrieval, tag-based image retrieval and tag expansion.}

During the discovery process, better described  in the remainder of the paper, the device itself can send its VUIP (coinciding with the VUIP of the owner) to its first social neighbors in the SIoT, i.e., nodes with which it has already established at least one social object relationship. We assume that each device has such a capability of profiling the interests of its owner and creating the corresponding VUIP.

Obviously, during the whole process of browsing the SIoT, the Source's VUIP exchanged among the devices \textit{must remain anonymous}. 

To our purposes, we also define a new social relationship between devices, the \textit{Co-Interest Object Relationship (C-IOR)}.
Such a relationship is established between two devices when the VUIPs of their device owners are sufficiently similar, i.e., when the degree of similarity between the VUIPs exceeds a certain threshold. 


\subsection{Basic Mechanism for the C-IOR establishment}
\label{subsec:D}

\begin{figure*}[t]
\centering
\includegraphics[scale=0.5]{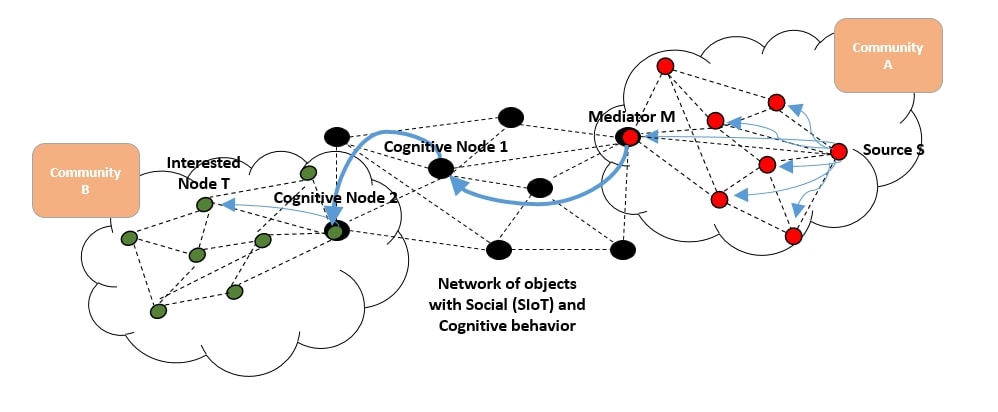}
\caption{Basic Mechanism for the establishment of the C-IOR.} 
\label{fig:meccanismobase}
\end{figure*}

In general, an Interest Neighbor of a $S$ device is a device connected via C-IOR to $S$, therefore it is a device sharing one or more interests with $S$.
We describe in the following the basic mechanism that characterizes the phases of Interest Neighbor discovery and corresponding C-IOR establishment. For greater clarity we will refer to Fig. \Ref{fig:meccanismobase}.

The $S$ device derives its owner's VUIP. 
$S$ sends its VUIP to all its first social neighbors. The first social neighbors of $S$ are the nodes with which $S$ has already established at least one SIoT relationship (such as C-LOR, SOR, etc.); among them there can be also cognitive objects, including the Mediator Object $M$. 
We assume that all the devices are cognitive. The cognitive objects can disclose the $S$'s VUIP also to cognitive objects belonging to other communities. 
Every first social neighbor of $S$ who receives the VUIP of $S$, including $M$, checks whether it is possible for it to establish a C-IOR with $S$ (based on the degree of similarity between VUIPs) and sends in turn the $S$'s VUIP (appropriately anonymized) to its own first social neighbors who have not  received the $S$'s VUIP yet.

The VUIP of $S$ is recursively delivered with a maximum Time to Live (TTL) of 6 hops (small world network property) set by the Source and decremented hop-by-hop.

If an interested node in another community, e.g., the $T$ node in Community B, based on the similarity between its own VUIP and the received S's VUIP, decides to establish a C-IOR with the owner of the received VUIP (it does not know that $S$ is the owner), then the C-IOR establishment request of $T$ (which includes $T$'s identity) goes backwards, forwarded by the intermediate nodes, until it gets to Source $S$. 
The reason why $T$ cannot directly send a C-IOR establishment request to $S$ but rather the establishment request of  $T$ is brought back to the Source by leveraging the intermediate nodes is that the $VUIP$ owner (i.e. $S$)  must always remain anonymous in order not to infringe $S$'s privacy.

In short, to ensure the privacy of the Source,  each intermediary knows only the identity of the "previous node" that forwarded the VUIP to it, and to which it will have to forward any request of C-IOR establishment coming from the interested node $T$.
After 6 hops, the VUIP expires and it is no longer forwarded.
Once the C-IOR establishment request of the concerned node reaches the Source, a C-IOR can be established that directly binds node $T$ and node $S$.

The Cognitive devices, the Mediators in particular, by mediating the propagation of the VUIP (Interest Descriptor) of $S$ from one community to another, and by enabling the establishment of the C-IOR between nodes belonging to different communities, allow the propagation of content across separate communities.


The  described mechanism, based on the VUIP propagation, is obviously general and can be implemented  with any type of Interest Descriptor obtainable for a given user.


\subsection{Enhanced  Discovery and Enhanced Diffusion}

The foundation of the Solid's ``Contacts" is the ``Contacts of Contacts" mechanism, in line with the more known ``Friends of Friends" mechanism. Accordingly, each node performing Discovery (search) can scan, in addition to the Contacts (friends) contained in its User POD, also the Contacts of its own Contacts, which have granted it authorization to access its Users' PODs. This is possible thanks to the link-following SPARQL 
\cite{sambra2016solid, mansour2016demonstration}.

We are assuming that in the User POD each user keeps her own list of contacts (\textit{Contact List}). 
In addition a user can scan the Contacts, contained in the User PODs of other users (her contacts), to which it is authorized to access.  

On the other hand, the list of nodes,  which each device is linked  through SIoT relationships (\textit{SIoT Contact List}) is locally stored  on the device itself, and specifies the type(s) of relationship(s) through which the nodes
are mutually linked. More details on how to create a distributed SIoT network are illustrated in the literature \cite{girau2016lysis}.

During the Discovery phase, a device will scan: (i) its User POD (i.e., its own Contact List), (ii) the User PODs of its Contacts that authorize it to access, (iii) and the User PODs of the Contacts reachable recursively (through the chain of authorizations) that authorize it to access.
In addition, the device will know its own SIoT Contact List.
Specifically,  the node will scan both the interested nodes and the nodes not interested in the content. 
The result of the Discovery will return only the interested nodes among the scanned ones.

For brevity we call them ED-IRC (Enhanced Direct Interested Reachable Contacts). 
We use the term ``Enhanced" when we leverage also the SIoT Contacts.

In short, with \textit{Direct} we identify nodes reachable in the Discovery phase, therefore directly reachable by the node that wants to disclose the content, while with \textit{Indirect} we mean the nodes that cannot be reached in the Discovery phase, but are anyway reachable in the Diffusion phase 
through intermediate nodes, which do not authorize (in the Discovery phase) the Source node to access their User POD.


After each node has performed Discovery and knows its ED-IRC, only Interested Nodes will be considered during the Diffusion phase. 
A node that wants to disclose content can disclose it directly to its ED-IRC.




\section{Performance Assessment}
\label{sec:6}

\subsection{ Generation of realistic SIoT datasets}
\label{sec:5}

    
We followed the procedure used in  \cite{njoo2018distinguishing} to obtain a reliable dataset on the meetings (co-locations) that take place between people. Accordingly, by starting from the Check-ins Dataset (Brightkite \cite{cho2011friendship}) we obtain a co-location if and only if two Check-ins 
of two different users took place within 250 meters and within 1800 seconds. 

While the mentioned procedure refers to contacts between persons, we are interested in the contacts between devices that may bring to the triggering of social relationships between them. Therefore, without losing generality, we assume that each person takes a mobile device with her and leaves a fixed device at home (her Home-Point). The assignment of a given model to a device (useful for establishing POR relationships) is carried out based on the ownership report of the Global Web Index in 2017 \cite{gwi2017} calculated on 50000 users. 

In this way, by replacing each user with their own devices, we can easily obtain  the meetings that took place between devices, from those that took place between their owners. Like in the aforementioned paper \cite{njoo2018distinguishing}, we will only consider users with at least 10 Check-ins and with at least 10 different Check-ins places. This allows us to exclude less active and therefore less interesting users.
        
Various methods are proposed in the literature to derive the positions of the \textit{Home-Points} of the users \cite{cho2011friendship}, \cite{scellato2011socio}, \cite{cheng2011exploring}. We recall here that the Home-Points are fundamental to derive the position of the fixed devices that are part of the SIoT network. The method we used to calculate the users' Home-Points is the one proposed in \cite{cheng2011exploring}. 

In general, there are places that we  call for convenience \textit{Points (or Places) of Interest} (PoI), towards which nodes will be more likely to go. We can see them as places of particular importance, places where specific activities are carried out, where people cultivate certain interests. Around these places we will have a greater concentration of meetings (in the reminder also referred to as co-locations). 

Each meeting will take place more or less near a PoI. Thus, we  assign a specific PoI to each meeting. 
This means that if a meeting took place near a certain PoI, with a certain probability the user (and her relevant device) has the Interest associated to that specific place. In particular, to assign an Interest to a co-location, we used a Foursquare dataset \cite{yang2016participatory, yang2015nationtelescope, yang2019revisiting}  that associates each PoI (in terms of latitude and longitude) with an Interest. In practice, by putting into relationship the meeting position and the PoI position in the Foursquare dataset, 
it is possible to assign a relevant Interest to each co-location. 

As for the Interests considered in our experiments, we started with the Foursquare Interests and we had to group them into Macro-categories, because they were defined in a very specific way. This had the effect that a large enough number of Communities with common Interest were not created to guarantee us a good statistical confidence in the results of our analyzes.

Each Foursquare Interest is described by a single keyword, while a Macro-category (Interest) is made up of a set of keywords. The interest associated with a user (or device) and also with a content will be a Macro-category (Interest). All Foursquare's Interests fall into 52 Macro-categories that we call Interests.  As an example, the Macro-categories we used in the performance evaluation studies illustrated in the remainder of the Section are: 
Sweet Food (Interest 3) including the Foursquare Interests: \{'Pastelaria', 'Ice Cream', 'Yogurt', 'Donut', 'Dessert'\}; Italian Food (Interest 4) including the Foursquare Interests: \{'Meatball', 'Wine', 'Pizza', 'Ice Cream'\};  Café Bar (Interest 6) including the Foursquare Interests: \{'Bistro', 'Breakfast', 'Cafe', 'Tea Room', 'Donut', 'Dive Bar', 'Cupcake', 'Coffee', 'Bar'\}.
    
 
Obviously, a meeting  near a certain PoI could happen casually. To understand if the user assiduously attends that PoI, a given  number of meetings must take place near that place, or better, \textit{near places of that type}. Therefore, we set a threshold on the minimum number of meetings near a PoI (set to 10 in the shown performance campaign) required  to assign that Interest to the user. In this way, from the co-locations (meetings) it is possible to obtain the PoI frequented by people, from which we can obtain their \textit{interests}. Each device is associated with an Interest Descriptor, i.e. a vector of words (keywords) that describe the interests of its owner.
    
Following the described procedure  we are able to establish the \textit{Communities}. In particular, from the Brightkite Dataset \cite{cho2011friendship} 
we get the \textit{Friendships} between people and we establish the \textit{Authorizations to Access Contacts} in Solid PODs.
In this way, assuming that there is a Source that wants to spread content related to a given Interest, we are able to obtain the \textit{Community}. 

At the same time, from the co-location events (meetings) we establish the \textit{SIoT relationships} according to the SIoT rules  available from the literature \cite{atzori2012social}.

\subsection{Use cases}
In the studies presented in the remainder of the paper, the objective is to compare the mean IRN percentage (ie, the percentage of interested nodes reachable) obtained by using the \textit{Enhanced SIoT} mechanism, which leverages the basic mechanism for the establishment of C-IOR, with the one obtained by the \textit{Friendships} mechanism, in which only Brightkite friendships are leveraged.

More specifically, in the  Friendships case, each node will be able to diffuse the Source content to all the interested nodes contained in its own Contacts List, in the Contact Lists of other users (her contacts) which it is authorized to access. In the Enhanced SIoT case, each node will be able to diffuse the Source content to the same interested node of the previous case, with the addition of those contained in its own SIoT Contact List.

A sufficient number of simulations were performed to be able to obtain statistical confidence in the IRN values shown in the curves.

\subsection{Assumptions}

The following assumptions hold:
\begin{itemize}
    \item All SIoT relationships are considered except the C-WOR, which as demonstrated in the article \cite{marche2017navigability} has a negligible contribution in terms of navigability \cite{marche2018dataset}.
    \item A threshold of 10 Check-ins is set in a specific type of PoI for the assignment of the relevant interests of a user.
    \item Each person brings a mobile device with her and leaves a fixed device at her home.
    \item In the Enhanced SIoT case (with C-IOR Basic Mechanism described in Sec. \ref{subsec:D}) node $A$ spreads the Source data to node $B$ of another community if and only if:
    \begin{itemize}
        \item the two nodes are connected via a SIoT relationship or via a SIoT relationships path (connection in the social graph of devices);
        \item the VUIPs of the two nodes have a similarity higher than a certain threshold (Cosine Similarity $\geq$ 0.5 \cite{zhou2016applying}). 
        The first two conditions imply the establishment of a C-IOR between the two nodes;
        \item the $B$ node has the specific interest of the data (which the $A$ node wants to spread) in its own VUIP. The third condition implies the presence of a C-IOR between the two nodes associated to such specific interest. 
    \end {itemize}
    
    \item Each node that has SIoT relationships with nodes belonging to other communities (communities other than its own) acts as a potential Mediator.
    
    \item Scenarios as realistic as possible are considered. A limit is set on the number of hops for the diffusion of the Source's VUIP (TTL) in the Discovery phase, as it is more realistic to assume that not all nodes are willing to spread the VUIP on behalf of another node. The percentage of nodes that spreads the Source VUIP to the different hops is varied during the simulations. In addition, since it is objectively less likely that a node makes its contacts available when increasing the social distance, then the percentage of nodes that provide authorization to access their PODs is assumed lower as the number of hops increases. 
    
    \item It is assumed that every Source that spreads its own content will spread it to all possible interested nodes. In particular, in the Friendships case it will spread it to all the interested nodes contained in its own Contacts List and in the Contact Lists of other users (her contacts) which it is authorized to access. In the Enhanced SIoT case interested nodes in its own SIoT Contact List are also considered. 
    
    \item Without losing generality, we assume that unless otherwise indicated we consider Interest 3 (``Sweet Food") and the related Communities.
    
    \item It is assumed that not only  interested nodes, but also not interested ones can authorize access to their PODs.  
    
    \item Unless otherwise indicated, all nodes, including isolated nodes, are considered.
    
\end {itemize}

\subsection{Performance by varying the number of nodes that spread the Source's VUIP}

The aim of the first performance evaluation is to investigate how the mean IRN percentage varies when varying the percentage of the nodes that diffuse the VUIP of the Source at each hop, by keeping fixed the percentage of nodes that authorize access to their PODs.
The nodes that spread the Source VUIP are the nodes that act as intermediaries, allowing the Source to reach Contacts otherwise unreachable. 
The reported results consider a percentage of the nodes that spreads the Source's VUIP at each hop equal to 100\%, 90\%, 60\%, 30\%, and 10\%, 
and a number of hops for the VUIP diffusion equal to 4.
All simulations were carried out in order to obtain a high statistical confidence (95\%). 

\begin{figure}[h]
\centering
\includegraphics[scale=0.27]{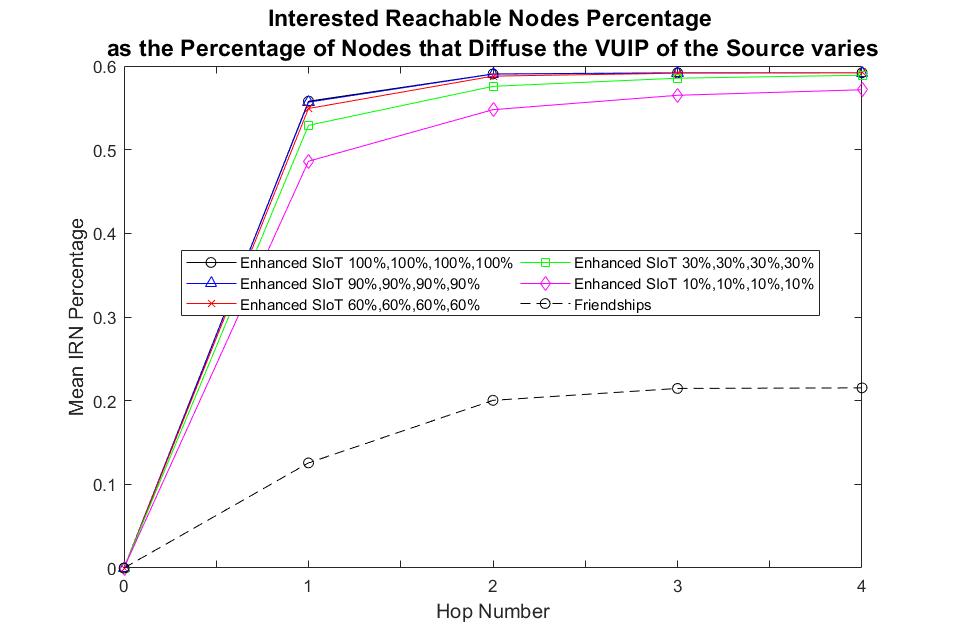}
\caption{Mean IRN percentage as the percentage of nodes that diffuses the Source VUIP at the different hops varies (Enhanced SIoT case vs. Friendships case.}
\label{fig:fig2}
\end{figure}

In Figure \ref{fig:fig2} the solid curves represent the trends obtained when exploiting all the social object relationships in the Enhanced SIoT case. The dotted curves represent the trends obtained if only Brightkite friendships are used (Friendships case). It is assumed that the percentage of nodes that authorize access to their POD is 100\% at the first hop.

By observing Figure \ref{fig:fig2} we can appreciate the higher values in terms of mean IRN percentage obtained in the Enhanced SIoT case compared to the Friendships case. This means that through the Enhanced SIoT it is possible to reach a greater number of interested nodes. This is due to the presence of SIoT relationships and of all the additional proposed features and mechanism previously described, from the Mediator object to the basic establishment mechanism for the C-IOR.  

The first two hops are those that have a greater increase in terms of mean IRN percentage (greater slope). We can note also the faster convergence in the Enhanced SIoT case compared to the Friendships case. This does not only mean that through the Enhanced SIoT a greater number of interested nodes can be reached, but also that they can be reached in a lower number of hops.

By observing Figure \ref{fig:fig2} it also clearly emerges, as we expected, that the obtained values in terms of mean IRN percentage increase with the increase in the percentage of nodes that diffuse the Source's VUIP, and with the increase in the number of hops. We can note that also in the worst Enhanced SIoT case (in which only  10\% of the nodes diffuse the Source's VUIP at each hop),  higher performance levels are achieved with respect to the Friendships case.


The low values obtained in general depend on the high number of interested isolated nodes present in the network for the specific scenario. As the number of hops increases, the increase in terms of mean IRN percentage becomes smaller, because most of the interested nodes that can be reached have already been reached.

\subsection{Performance as the percentage of nodes that authorize access to their PODs changes}

\begin{figure}[h]
\centering
\includegraphics[scale=0.28]{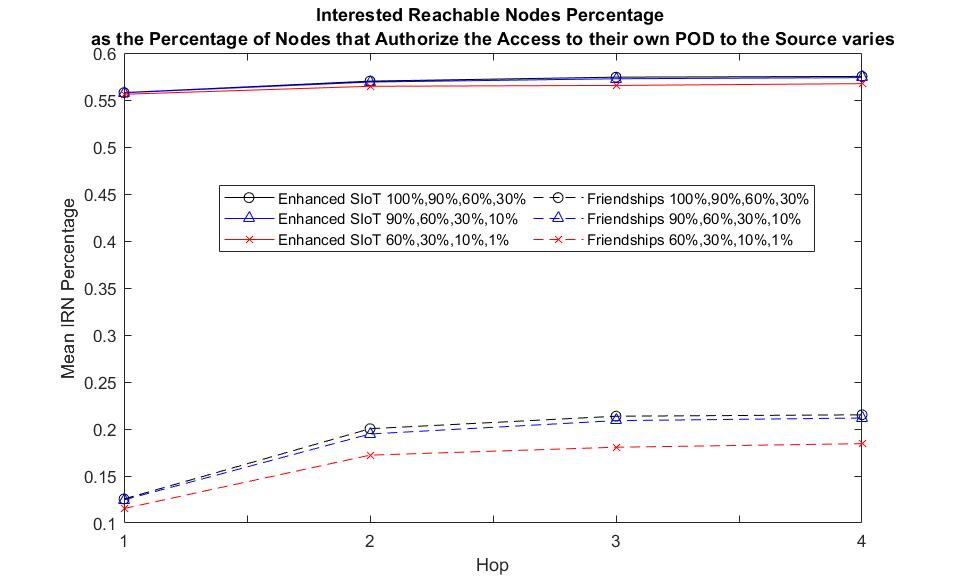}
\caption{Mean IRN percentage when varying the percentage of nodes authorizing access to their POD at the different hops (Enhanced SIoT case vs. Friendships case.}
\label{fig:fig4}
\end{figure}

The second study aims to investigate how the mean IRN percentage varies with the percentage of nodes that authorize access to their PODs at different hops, by keeping the percentage of nodes that spread the Source's VUIP fixed.
Let us consider the limit of 4 hops, in which there will be nodes authorizing the access to their PODs. The labels of Figure \ref{fig:fig4} report the percentages of nodes that authorize the Source to access their PODs in each of the 4 hops. 
\textcolor{black}{The reason why in Figure 3 we have set decreasing percentages of nodes that authorize the Source to access their PODs at each hop is that it is correct to assume that friends are more willing to authorize access to their PODs than friends of friends and so on. According to the real social dynamics in networks, the more socially distant one node is from another one, the less likely this node will authorize this latter node to access its POD. }

Again, in Figure \ref{fig:fig4} the solid curves represent the trends obtained when all the social object relationships are considered in the Enhanced SIoT case. The dotted curves represent the trends obtained if only Brightkite friendships are used (Friendships case). We assume that the percentage of nodes diffusing the Source VUIP is 100\% at the first hop, i.e. all the nodes spread the Source's VUIP.

From Figure \ref{fig:fig4} we can note the higher  values in terms of mean IRN percentage obtained in the Enhanced SIoT case compared to the Friendships case. Also here, the Enhanced SIoT is able to reach a greater number of interested nodes.  
By observing Figure \ref{fig:fig4} it also clearly emerges, as expected, that the obtained values in terms of mean IRN percentage increase with the increase in the percentage of nodes that authorize the access to their PODs, and with the increase in the number of hops (in which there are node that provide authorization to access their PODs to the Source). We can note that also in the worst Enhanced SIoT case,  higher performance levels are obtained with respect to the Friendships case.
Here, again the low values in general depend on the high number of interested isolated nodes present in the network.
The reader can note that the gain obtained with a higher percentage of nodes, which authorize to access their contacts, is more accentuated in the Friendships case than in the Enhanced SIoT case. Also, the first two hops are those that show a greater increase in terms of mean IRN percentage (greater slope of the curves). This is due both to the fact that with the increase in the number of hops, most of the nodes that can be reached have already been reached, and to the fact that in the first hops we set higher percentages of nodes authorizing access to their PODs. This latter assumption has not to surprise, because it is correct to assume that friends are more willing to authorize access to their PODs than friends of friends and so on. The more socially distant a node is, the less likely this node will authorize access to its POD.

\subsection{Performance by varying the kind of SIoT relationships between devices}

\begin{figure}[hp]
\centering
\includegraphics[scale=0.28]{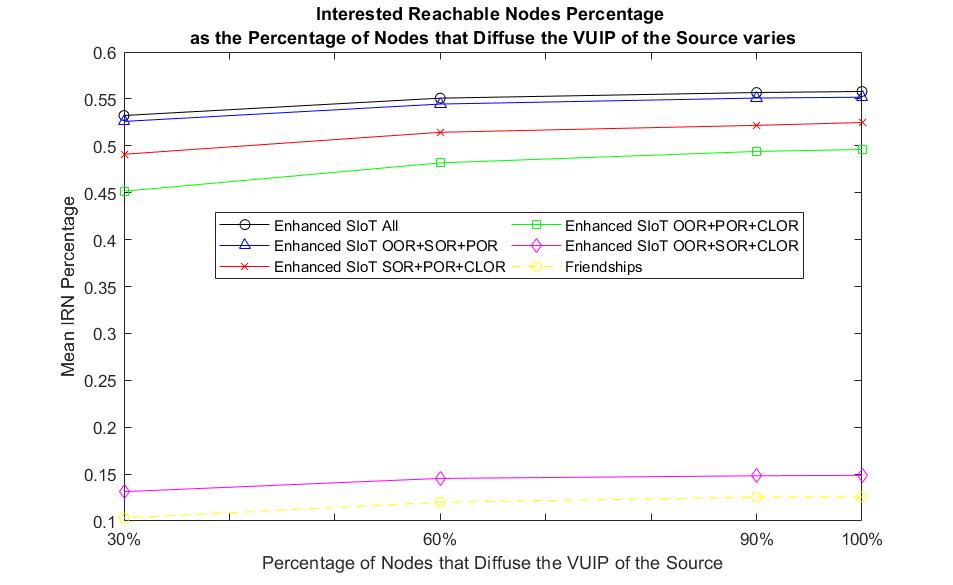}
\caption{Mean IRN percentage for different combination of SIoT relationships, as the percentage of nodes that diffuses the Source VUIP changes (Enhanced SIoT vs. Friendship).}
\label{fig:fig7}
\end{figure}

A further objective of our study is to observe how the mean IRN percentage changes when the combination of SIoT relationships vary. For this purpose, simulations have been conducted in which six different combinations of SIoT relationships are considered. 
Figure \ref{fig:fig7} shows the variation of the mean IRN percentage, assuming that the 100\%, 90\%, 60\%, and 30\% of nodes respectively spreads the VUIP of the Source (act as intermediaries), in the Enhanced SIoT case.

A first evident result is that POR is clearly the social object relationship that weighs most on the obtainable mean IRN percentage values, followed by the SOR, the OOR, and the C-LOR. POR friendships in fact depend only on the model of the device and are often relationships that connect devices that are very distant from each other and belong to different communities. Given their characteristic of being "long-range" relationships, the relevant role, confirmed by the curves, in connecting users belonging to different communities otherwise separated was expected.
The advantage in terms of the considered metric that the Enhanced SIoT case offer compared to the Friendships case, for any combination of SIoT relationships, is evident from the curves shown in Fig. \ref{fig:fig7}; the values in terms of mean IRN percentage obviously increases with the increase in the percentage of nodes that spread the Source VUIP.

\subsection{Performance by varying the type of Interest}

Up till now, in our performance evaluation study we have always considered Interest 3. Obviously, the performance figures may depend on the choice of the interest and it is important to understand how the scenario resulting from a change in the interest influences the performance.  

\textcolor{black}{Therefore, a first measurement campaign was aimed at analyzing in which scenarios (each characterized by a different Interest) there is a greater number of nodes belonging to the Giant Component, and what is the increase of this component when passing from the Friendships case to that Enhanced SIoT. The percentage of nodes belonging to the Giant Component is important because it tells us what is the largest subset of nodes that are connected to each other.
Here we consider six hops for the diffusion of the Source VUIP and we establish that the percentage of nodes that spread the VUIP of the Source and that authorize access to their PODs at the Source are both 100\% at each hop. }

\begin{figure}[h]
\centering
\includegraphics[scale=0.28]{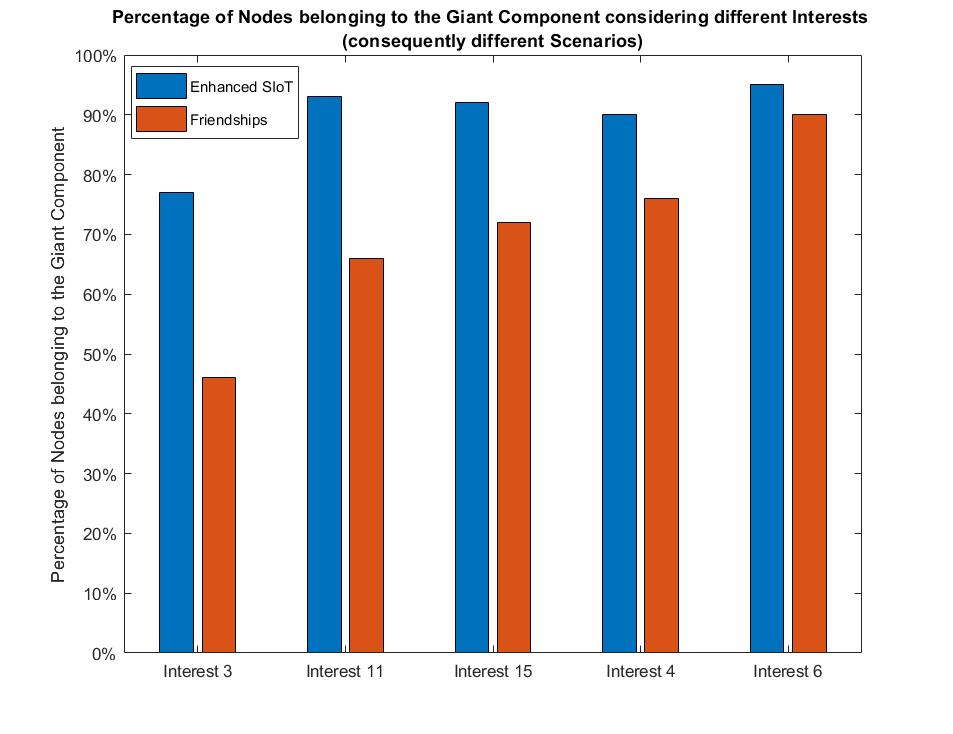}
\caption{\textcolor{black}{Percentage of nodes belonging to the Giant Component considering different Interests.}} 
\label{fig:fig11}
\end{figure}

\textcolor{black}{Figure \ref{fig:fig11} reports the percentage of nodes belonging to the Giant Component, out of the total number of nodes of the Scenario, in the case of Friendships. On the x-axis, the Interests have been sorted by increasing values of nodes belonging to the Giant Component, in the case of Friendships. The percentage of nodes belonging to the Giant Component in the Enhanced SIoT case is also shown in blue.}

\begin{figure}[h]
\centering
\includegraphics[scale=0.28]{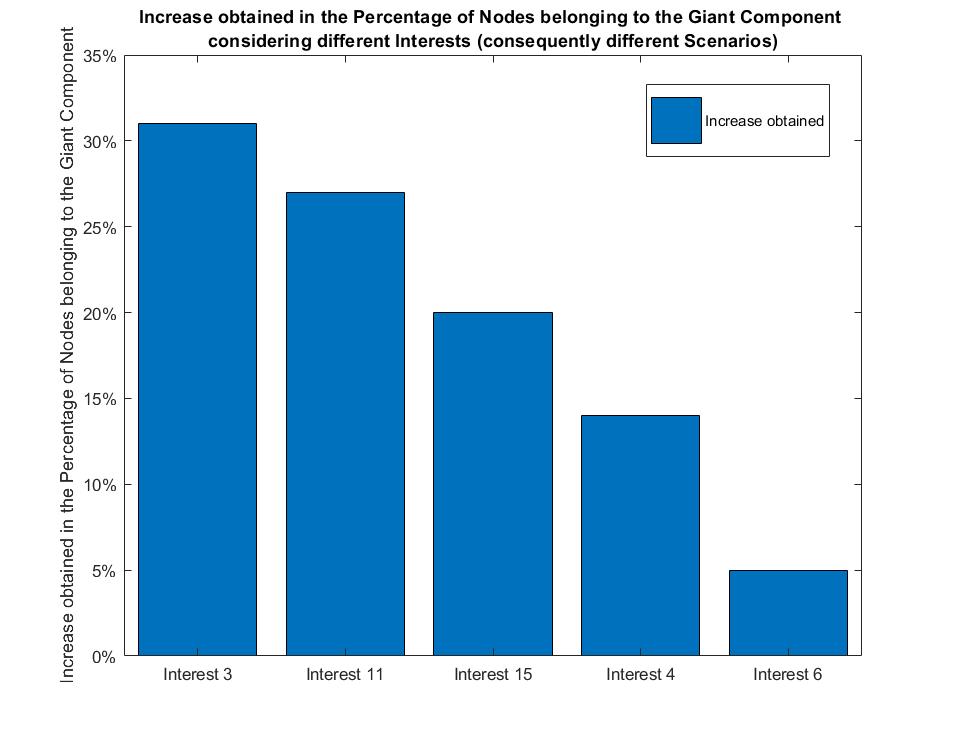}
\caption{\textcolor{black}{Increase obtained in the percentage of nodes belonging to the Giant Component considering different Interests.}} 
\label{fig:fig12}
\end{figure}

\textcolor{black}{As the Interest varies, we obtain different scenarios/graphs by considering, for each case, the nodes that possess that specific Interest. Considering only the Friendships as arcs of the graph (Friendships case, red labels) we can obtain graphs in which the nodes are almost all connected to each other (many nodes belong to the Giant Component), as in the Interest 6 case, or in which they are poorly connected, as in the Interest 3 case.}

\textcolor{black}{By observing Figure \ref{fig:fig11} it can be seen that the lower the number of nodes belonging to the Giant Component in the Friendships case, the greater the increase of this value which is obtained in the Enhanced SIoT case (for ease of reading, Figure \ref{fig:fig12} shows such an increase).
This means that the worse the starting scenario (in the case of Friendships), the greater the gain achieved with the Enhanced SIoT. 
The difference in the results obtained is due to the fact that, as expected, in poorly connected Scenarios in the Friendships case, the established Social Object Relationships are able to connect a larger number of nodes that were not already connected by Friendships.}

\textcolor{black}{We need now to better understand which is the increase in the percentage of IRN nodes, which play an important role in our distributed social network. Without losing generality, we focus on Interests 3, 4, and 6, and again consider six hops for the diffusion of the Source VUIP, and 100\% of nodes that, at each hop, spread the Source VUIP and authorize access to their PODs. }

\begin{figure}[h]
\centering
\includegraphics[scale=0.28]{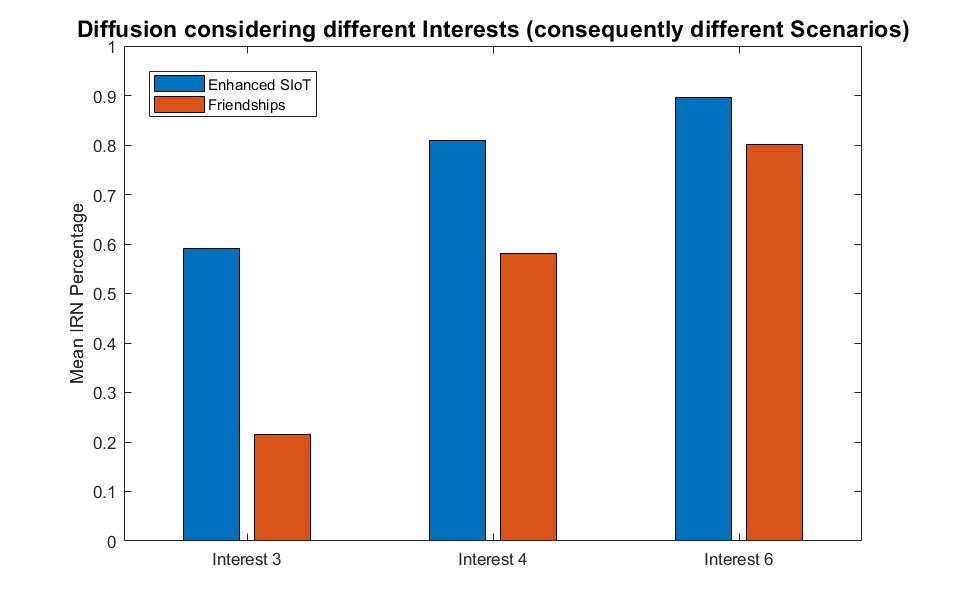}
\caption {Mean IRN percentage when varying the considered Interest.}
\label{fig:fig8}
\end{figure}

\textcolor{black}{In Figure \Ref{fig:fig8} we can see that in the graph obtained considering Interest 6, already well connected in the Friendships case, the obtained Mean IRN percentage value is very high. On the contrary considering Interest 3, with only a few nodes belonging to the Giant Component in the Friendships case, the obtained Mean IRN percentage value is very low.}

\begin{figure}[h]
\centering
\includegraphics[scale=0.28]{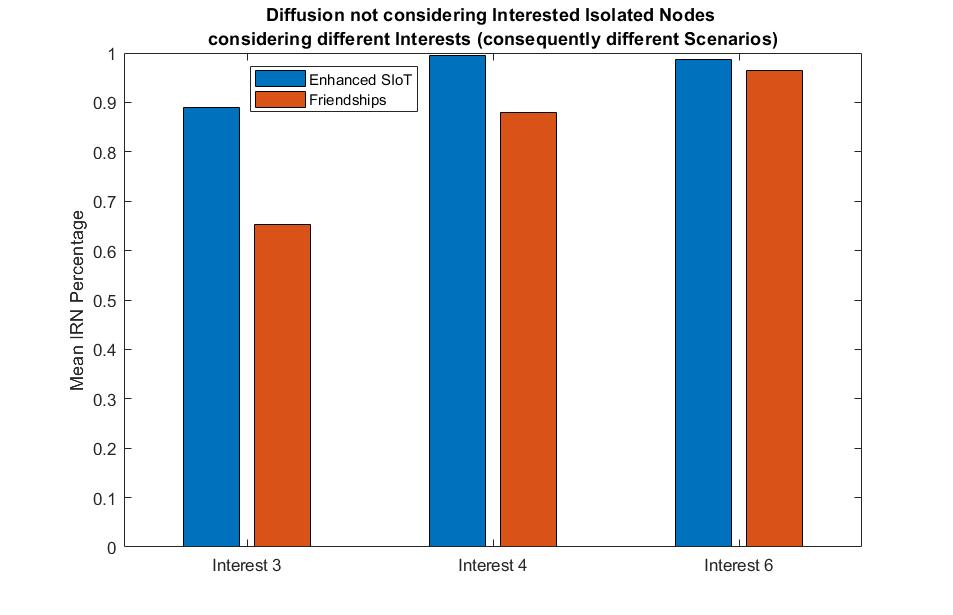}
\caption{Mean IRN percentage when considering different Interests (isolated nodes NOT considered).} 
\label{fig:fig10}
\end{figure}

\textcolor{black}{The much more important advantage obtained by using the Social Object Relationships are confirmed in the Interest 3 case, wherein it allows to greatly improve the Mean IRN percentage value.
On the contrary, using the Social Object Relationships is of little relevance in the Interest 6 case, wherein the advantage obtained is small since high values of Mean IRN are attainable without its introduction.}
\textcolor{black}{Like before, if we do not consider the interested but isolated nodes, the behaviours remain the same while the reachable performance levels are higher, as shown in Figure \ref{fig:fig10}}

\begin{figure}[h]
\centering
\includegraphics[scale=0.28]{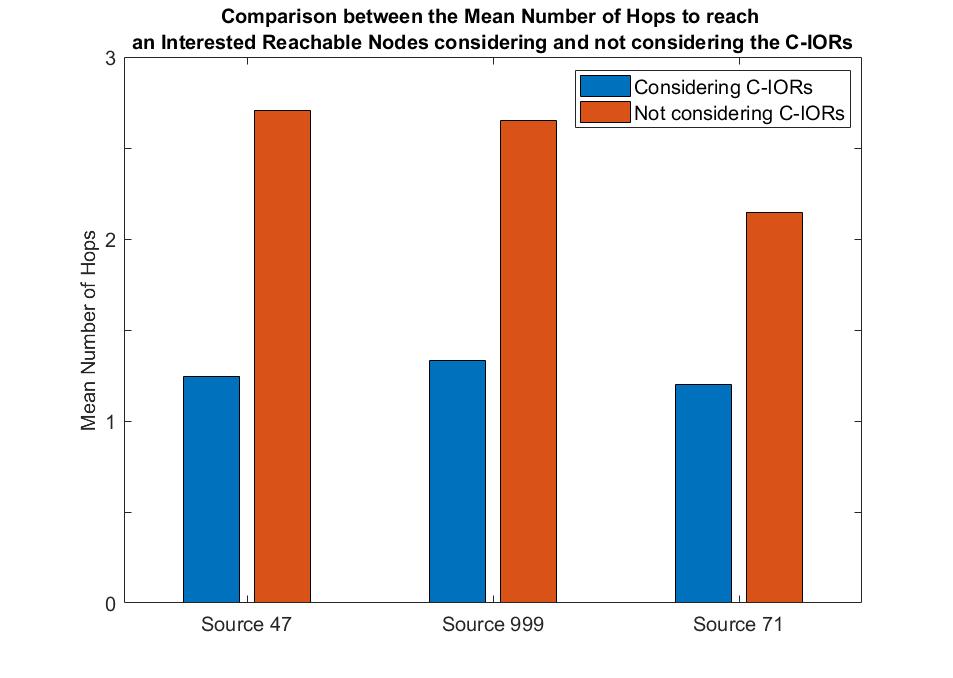}
\caption {Comparison between the Mean Number of Hops employed by the Source to reach an Interested Reachable Node when considering/not considering the C-IORs.}
\label{fig:fig20}
\end{figure}

The last study conducted aimed at demonstrating and quantifying the advantage deriving from the adoption of the C-IOR object friendships in terms of mean reduction of the number of hops employed by the Source to reach all the interested nodes during the Discovery phases. A faster discovery enabled by the use of C-IORs has been confirmed, as in our experiments we have always found for each source an average number of hops that is almost halved compared to the case in which C-IORs are not used (an example referring to  three  nodes randomly chosen is depicted in Figure \ref{fig:fig20}, but a similar behaviour is found for all nodes in the considered population).  
The manifest advantage in terms of delay reduction during Discovery is only paid in terms of a slight increase in the computational complexity introduced by the basic mechanism of the C-IOR establishment.

\subsection{Final remarks}




In conclusion, in all the conducted studies the advantage achieved in the Enhanced SIoT case compared to the Friendships case is evident, thanks to the possibility of using the SIoT relationships. Furthermore, the contribution given by the C-IOR relationship appears to be significant. 
The reason for this is the fact that the basic mechanism of C-IOR allows the establishment of direct social links (C-IOR) to interested nodes that would otherwise be connected only through a chain of SIoT relationships that might also involve nodes not interested to the content.


\textcolor{black}{As for the use of hub nodes as preferential relaying nodes in order to optimize the performance, we have to clarify that for the purposes of our work, the number of links in the social graph of a given node is not so important, but rather the nature of these links (i.e., the nature of the SIoT friendship that the link represents) is important. In fact, a hub node with a very high degree of connection but only with links to other nodes of its own community of interest is much less attractive, for the purposes of its mediator role, than a node that is not a hub and has just a few SIoT links but with objects from other communities with similar interests.} 

\textcolor{black}{In our work we have assumed the Source S node sending its VUIP to all its first social neighbors (nodes with which S has already established at least one Social Object Relationship) and so on recursively. Furthermore, among the Social Object Relationships it has been noted that POR links (long range) are particularly important to connect different communities that share the same interest.
In our simulations, as we are dealing with networks that are not too large in terms of the number of nodes and arcs, we have chosen to consider when possible all the nodes (except in cases with percentages other than 100\%).}

\textcolor{black}{In real cases, obviously, to avoid spreading to all nodes, relying on hub nodes (intended, however, as nodes with a large number of POR relationships) could certainly bring advantages. This could be a starting point for a later work.}

\textcolor{black}{A final point that we intend to highlight relates to possible extensions of the concepts introduced in this paper also to environments other than those of traditional digital social networks, which could be the subject of future investigations. Certainly, a very promising environment in which to test the potential of the new concepts proposed is that of a Vehicular Social Network (VSN). In fact, VSNs are intrinsically characterized by a decentralized nature and present a scenario with cars playing the role of "content prosumers", i.e. both producers and consumers, as it is understood in our work. An aspect that differentiates VSNs from the OSNs we have considered in this work is the \textit{highly dynamic} nature of VSNs, in which social links are built ``on-the-fly" and have a short lifetime, whenever the community members become neighbors to each other \cite{Vegni}, \cite{Rahim}. More specifically, a vehicle can enter a social network and stay for a limited time by exchanging messages with the neighboring vehicles about a given topic related to the same interest or experiences, and it can participate to a known social network whenever approaching a specific area of interest \cite{Vegni}, \cite{Rahim}.
Given there differences, the result of our research cannot be applied ``as it is" to VSN environments, since we aim to increase the level of connectivity between communities that belong to a distributed human OSN but which are defined via fairly stable links, through the use of SIoT, which is also decentralized and  based on stable inter-device links. Besides, there would be a need to include concepts that are consistent with the destructive paradigm introduced by SOLID also in the VSN environment.}

\textcolor{black}{It can be concluded that there is a need for some adaptations to extend what is described in the paper to be applied to vehicular scenarios. There are however good possibilities to achieve this, if we consider that the concept of clustering a VSN into social groups composed of vehicles with common interests is already widely spread, and that the SIoT principles have also been integrated with IoT features in the vehicular environment \cite{Alam}, \cite{Floris}.}

\section{Conclusions}
\label{sec:conclusions}

In this paper we have proposed a new platform model  for DOSNs based on the joint use of the Solid platform and the new paradigm of \ac{SIoT}, emerging with increasing strength.
Evidence has been provided of the fact that by coupling these two concepts together it is possible to arrive at the design of a modern \ac{DOSN} platform that permits users to maintain control over their personal information and, at the same time, effectively limits the intrinsic drawbacks that in the past made \acp{DOSN} unattractive compared to centralized solutions.
Through a simulation campaign aimed at comparing the ability to connect users with the same interest but belonging to separate communities within a \ac{DOSN} platform, it was possible to prove that the road traced has the potential to make distributed social networks more attractive and to facilitate their large-scale deployment. This can be achieved thanks to the synergies that can be obtained between human users and social devices.

\end{document}